\documentclass[letterpaper, 10pt, conference]{ieeeconf}

\usepackage{amsmath,amssymb,euscript,yfonts,psfrag,latexsym,dsfont,graphicx}
\usepackage{bbm,color,amstext,wasysym,flushend,subfig,parskip,balance}
\usepackage[ruled,vlined]{algorithm2e}
\usepackage{algorithmic}
\usepackage{float}
\usepackage{hyperref}
\graphicspath{{./},{./figures/}}

\newcommand{\mR}{{\mathbb R}}

\newcommand{\cN}{{\mathcal N}}

\definecolor{grey}{rgb}{0.6,0.6,0.6}
\definecolor{lightgray}{rgb}{0.97,.99,0.99}

\IEEEoverridecommandlockouts
\title{\LARGE \bf Stein particle filtering }

\author{Jiaojiao Fan, Amirhossein Taghvaei, and Yongxin Chen
\thanks{This work was supported by NSF under grant 1942523 and 2008513.}
\thanks{J. Fan and Y.\ Chen are with the School of Aerospace Engineering,
Georgia Institute of Technology, Atlanta, GA; {\{jiaojiaofan,yongchen\}@gatech.edu}}
\thanks{A. Taghvaei is with the Department of Mechanical and Aerospace Engineering,
University of
California, Irvine, CA; {ataghvae@uci.edu}}}

\begin{document}


\maketitle

\begin{abstract}
We present a new particle filtering algorithm for nonlinear systems in the discrete-time setting. Our algorithm is based on the Stein variational gradient descent (SVGD) framework, which is a general approach to sample from a target distribution. We merge the standard two-step paradigm in particle filtering into one step so that SVGD can be used. A distinguishing feature of the proposed algorithm is that, unlike most particle filtering methods, all the particles at any time step are equally weighted and thus no update on the weights is needed. We further extended our algorithm to allow for updating previous particles within a sliding window. This strategy may improve the reliability of the algorithm with respect to unexpected disturbance in the dynamics or outlier-measurements. The efficacy of the proposed algorithms is illustrated through several numerical examples in comparison with a standard particle filtering method.
\end{abstract}

\section{Introduction}


Particle filters are sequential Monte-Carlo based algorithms designed to numerically approximate the solution to the filtering problem~\cite{gordon1993novel,doucet2009tutorial}. The basic idea in particle filter is to approximate the posterior distribution with weighted empirical distribution of particles. The key step in approximation is importance sampling, where the weights of the particles are updated according to the Bayes rule that involves the likelihood of the observation signal. The issue with this approach is that after several time-steps, there are only a few particles with large wights, while most of the particles have nearly zero weights. This issue is referred to as particle degeneracy~\cite{gordon2004beyond}. To mitigate the issue, particle filters perform a resampling step, where new independent particles with uniform weights are sampled from the weighted empirical distribution. Theoretically, it is shown that the particle filters are exact as the number of particles grows to infinity, with error rate $O(N^{-\frac{1}{2}})$~\cite{del2001stability,cappe2006inference}. However, both empirically and theoretically, it was discovered that particle filters perform poorly in high-dimensional problems~\cite{bengtsson2008curse,beskos2014error,rebeschini2015can}.

In this work, we propose a particle filtering framework which doesn't suffer from particle degeneracy. Departing from most particle filtering methods which iteratively update the weights of the particles through important sampling, our approach iteratively update only the locations of the particles. Consequently, all the particles at each time step remain uniformly weighted throughout. Thus, the effective particle size remains high all the time, a desirable property for particle filtering, particularly for high-dimensional applications. 

Our particle filtering method is based on the Stein variational gradient descent (SVGD) framework. The latter is a general purpose sampling technique which can generate samples from any given probability density as Markov chain Monte Carlo (MCMC) methods. One major difference between SVGD and MCMC is that the samples generated using SVGD are deterministic. It was argued that SVGD is less vulnerable to mode collapse than MCMC \cite{LiuWan16}. Moreover, the SVGD only need to know the target probability density up to a normalization factor. Thus, it is suitable for sampling from posterior distribution where the normalization factor is not available. In our method, we approximate the posterior distribution up to a normalization factor of the next time step using the particles of the current step and the next measurement. With this approximated posterior distribution we can adopt SVGD to sample from it. Our particle filtering algorithm is an iterative implementation of SVGD in this manner.

A sliding window version of this algorithm is also developed. It iteratively samples the trajectories of the posterior dynamics over a fixed-size sliding window instead of one-time point. In this sliding window particle filtering algorithm, each measurement is used multiple times as each measurement belongs to several different sliding windows. This reuse of measurement is helpful to improve the tolerance of the algorithm to mistakes made at some time step. Another potential benefit of this method is higher robustness with respect to outlier observations. 

One standard particle filter is Sampling Importance Resampling particle filter (SIR), which has been shown to be stable and well-performed in various settings. 
There is a growing interest in developing nonlinear   filter algorithms that avoid the weight degeneracy issue in particle filters by replacing the importance sampling step with a control-based approach. There are identified under the category of controlled interacting particle systems. The earliest example is the ensemble Kalman filters~\cite{evensen1994sequential,bergemann2012ensemble} which provides an exact solution to the filtering problem in linear Gaussian setting and has been widely applied in meteorological applications.  Feedback particle filter~\cite{yang2013feedback,yang2016multivariable,taghvaei2020diffusion} provides generalization of the ensemble Kalman filter to nonlinear and non-Gaussian setting~\cite{taghvaei2018kalman}. Other examples of controlled interacting particle systems include Crisan and Xiong filter~\cite{crisan2005approximate}, particle flow filters~\cite{daum2010exact}, and ensemble transform particle filter~\cite{reich2013nonparametric}. 

The rest of the paper develops as follows. In Section \ref{sec:backg} we present the background on particle filtering and the Stein variational gradient descent method. The proposed particle filtering algorithms are presented in Section \ref{sec:stein}. It is followed by several numerical examples in Section \ref{sec:num} and a concluding remark in Section \ref{sec:conclusion}.

\section{Background} \label{sec:backg}

In this section,  we present the mathematical formulation of the nonlinear filtering problem in discrete-time setting, a brief description  the particle filter algorithm, and a brief introduction to the Stein variational gradient descent method.

\subsection{Nonlinear filtering problem}
Consider the following dynamical system 
\begin{subequations}\label{eq:dyn}
    \begin{eqnarray}\label{eq:dyn1}
        x_{t+1} &\sim& p(x_{t+1}\mid x_t),\quad x_1 \sim p(x_1),
        \\\label{eq:dyn2}
        z_t &\sim& p(z_t\mid x_t),
    \end{eqnarray}
\end{subequations}
for $t  \in \mathbb{N}$ where  $x_t \in \mR^d$  denotes the state of the system at time $t$, and $z_t \in \mR^l$ denotes the observation process. Here $p(x_{t+1}|x_t)$ represents the transition probability of the Markov process $x_t$ and $p(z_t|x_t)$ represents the likelihood function of observing $z_t$  given $x_t$. The objective of the filtering problem is to compute the posterior distribution $p(x_t|Z_t)$ where $Z_t:=\{z_1,\ldots,z_t\}$ denotes the history of observations up to time $t$.

The posterior distribution can be computed recursively in two steps, using the Markov property of $x_t$ and the Bayes rule. In particular, given $p(x_t|Z_t)$, the posterior distribution at the next time step $p(x_{t+1}|Z_{t+1})$ is given by:
\begin{subequations}\label{eq:one_step_dist}
\begin{align}
         p(x_{t+1}|Z_{t})
     & =\int_{x_t} p(x_{t+1}|x_{t})p(x_{t}|Z_t),\\
         p(x_{t+1}|Z_{t+1})
     & =\frac{p(z_{t+1}|x_{t+1}) p(x_{t+1}|Z_t)}{p(z_{t+1}|Z_t)}.
\end{align}
\end{subequations}
Although the posterior admits a recursive update law in principle, it is numerically challenging to implement the update law in practice due to the fact  the posterior distribution is infinite-dimensional and    computation of the normalization constant is challenging.

\subsection{SIR particle filter}
The main idea in particle filters is to approximate the posterior distribution with weighted empirical distribution of particles
\begin{equation}
    p(x_t|Z_t) \approx \sum_{i=1}^n w^i_t \delta_{x^i_t}
\end{equation}
such that $\sum_{i=1}^n w^i_t=1$, where $\delta_x$ is the Dirac delta distribution located at $x$. There are several standard variations of the particle filter that are being used in practice. Each variation corresponds to a different choice of proposal density to update the particle location and evaluate the importance sampling weights. In this paper, we consider the sequential importance resampling (SIR) particle filter where the particles are update according to the dynamic model leading to the following update law for particles and weights~\cite{arulampalam2002tutorial}   
\begin{align*}
    x^i_{t+1} &\sim p(x_{t+1}|x^i_t)\\
    w^i_{t+1} &= \frac{w^i_t p(z_{t+1}|x^i_{t+1})}{\sum_{j=1}^n w^j_t p(z_{t+1}|x^j_{t+1})}. 
\end{align*}
To avoid the weight degeneracy issue, a resampling step is carried out after several steps, where $n$ independent new samples are generated from the weighted distribution $\tilde{x}^i_t \sim \sum_{i=1}^n w^i_t \delta_{x^i_t}$. After the resampling procedure, the posterior distribution is approximated with empirical distribution of the new samples with equal weights, i.e. $p(x_t|Z_t) \approx \frac{1}{n}\sum_{i=1}^n \delta_{\tilde{x}^i_t}$.

\subsection{Stein variational gradient descent}
The Stein variational gradient descent (SVGD) \cite{LiuWan16} is a sample-based variational inference technique. It enables efficient sampling from a target probability density $q(x)$ by gradually transforming samples from an arbitrary initial distribution to samples of $q(x)$.

Let $\{x^i_\ell\}_{i=1}^n$ be the samples at the $\ell$-th iteration, then the SVGD algorithm evolves the particles with
\begin{equation}\label{eq:stein}
    x^i_{\ell+1} =  x^i_{\ell} + \epsilon_\ell \hat{\phi}^*(x^i_\ell),
\end{equation}
where
\begin{equation}\label{eq:stein_phi}
    \hat{\phi }^*(x)=\frac{1}{n} \sum_{j=1}^n \left[k(x^j_\ell, x)\nabla_{x^j_\ell} \log q(x^j_\ell) + \nabla_{x^j_\ell} k(x^j_\ell, x)\right].
\end{equation}
Here $k(\cdot,\cdot)$ is a positive definite kernel \cite{liu2016kernelized,chwialkowski2016kernel}, e.g., the Gaussian kernel $k(x,x')=\exp(-\frac{1}{h} \|x-x'\|_2^2)$.

\begin{algorithm}[tb]
    \caption{Stein variational gradient descent (SVGD)}
    \label{al:svgd}
    \begin{algorithmic}
        \STATE {\bfseries Input:} A target distribution $q(x)$ and the initialized particles $\{x_{0}^i\}_{i=1}^n$.
        \FOR{$\ell=0,\ldots,L-1$}
        \STATE { $ x_{\ell+1}^i=x_{\ell}^i + \epsilon_\ell \hat{\phi}^*(x_{\ell}^i)$,
        where $\hat{\phi }^*(x)$ equals to \\
        $\frac{1}{n} \sum_{j=1}^n \left[k(x^j_\ell, x)\nabla_{x^j_\ell} \log q(x^j_\ell) + \nabla_{x^j_\ell} k(x^j_\ell, x)\right].$
        }
        \ENDFOR
        \STATE {\bfseries Output:} A set of  particles $\{x_{L}^i\}_{i=1}^n$ that approximates the distribution $q(x)$.
    \end{algorithmic}
\end{algorithm}

Note that the update \eqref{eq:stein_phi} requires only querying $\nabla_x\log q$, thus the normalization constant in $q$ is not important. This property makes it ideal for Bayesian inference problems where the computation of the normalization constant is challenging. We also remark that, unlike the particles in SIR particle filter, which are accompanied by weights, the particles $\{x^i_\ell\}_{i=1}^n$ in SVGD are always equally weighted, thus, no update for weights is needed.

In the mean-field limit, the SVGD \eqref{eq:stein_phi} is the gradient flow of the KL-divergence between the particle distribution and the target distribution $q(x)$ over the velocity field $\hat\phi$ with respect to the norm in a reproducing kernel Hilbert space (RKHS) associated with the kernel $k(\cdot,\cdot)$. Thus, the convergence of SVGD is guaranteed in the mean-field limit. We refer the reader to \cite{LiuWan16} for more detailed discussion on SVGD.

\section{Stein particle filtering} \label{sec:stein}

In this section, we propose stein particle filtering algorithm to iteratively estimate the posterior distribution $p(x_t|Z_t)$. It replaces the important sampling step in traditional particle filtering such as SIR with SVGD. As a result, the proposed algorithm only updates the position of the particles as the weights remain uniform throughout.
We present two variations of the algorithm: Sequential Stein particle filtering and Sliding window Stein particle filtering. The former sequentially update the particles to approximate $p(x_{t+1}|Z_{t+1})$ based on the particles that approximate $p(x_t|Z_t)$ as in most particle filtering algorithms. It only updates the particles associated with posterior distribution at the current step. The latter updates the particles for previous steps as well up to a fixed-size time window.


\subsection{Sequential Stein particle filtering}
In the Sequential Stein particle filtering algorithm, the posterior distribution 
$p(x_t |Z_t)$ is approximated by empirical distribution of $n$ particles $\{x_t^i\}_{i=1}^n$ with equal weights according to $p(x_t|Z_t) \approx \frac{1}{n} \sum_{i=1}^n \delta_{x^i_t}$. For $t>1$,  upon using this approximation in Equation \eqref{eq:one_step_dist}, we have the  the following expression for the posterior at the next time step:
\begin{equation}
    \label{eq:stride1_numeric}
    p(x_{t+1} | Z_{t+1}) \propto \left[\frac{1}{n} \sum_{i=1}^n p(x_{t+1}|x_t^i)\right] p(z_{t+1}|x_{t+1}).
\end{equation}
The main idea of the proposed algorithm is to apply the SVGD algorithm (Algorithm \ref{al:svgd}) to sample $n$ particles $x^i_{t+1}$ from the posterior $p(x_{t+1} | Z_{t+1})$. Then, this new set of particles are used to approximate the posterior distribution $p(x_{t+1} | Z_{t+1})$. This approximation is then used for the sampling of the next step. For $t=1$, the particles $x^i_1$ are generated by implementing the SVGD algorithm to sample from $p(x_{1}|Z_{1})$ given by 
\begin{equation}\label{eq:one_step_p1}
    p(x_{1} | Z_{1}) \propto p(z_1|x_1)p(x_1).
\end{equation}

There are two approaches to initialize the particles for implementing the SVGD algorithm in the particle filtering problems:  the particles can be initialized from  $x^i_t$, or they can be sampled from the Markov transition probability  $p(x^i_{t+1}|x^i_t)$. Numerically, we observed that the  latter initialization is more effective as it gives samples that are closer to the desired posterior distribution.
We name this algorithm the Stein particle filtering algorithm and present the details in {Algorithm \ref{al:Iterative}}. 



\begin{algorithm}[tb]
    \caption{Sequential Stein particle filtering}
    \label{al:Iterative}
    \begin{algorithmic}
        \IF{$t=0$}
        \STATE {\bfseries Input:} initial dist. $p(x_1)$, observation $Z_1$
        \STATE {\bfseries Initialize:} $x^i \sim p(x_1), i=1,2, \ldots, n$
        \STATE {\bfseries Compute:} $\{\hat{x}^i\}=\text{SVGD}\left(p(x_{1}|Z_{1}),\{x^i \} \right) $ with Algorithm \ref{al:svgd}.
        \STATE {\bfseries Output: $\{x_{1}^i\}_{i=1}^n=\{\hat{x}^i\}_{i=1}^n$}
        \ELSIF{$t \geq 1$}
        \STATE {\bfseries Input:}
        $n$ particles from the last time step $\{x_{t}^i\}_{i=1}^n \sim p(x_t|Z_t)$, observations $Z_t$
        \STATE { \bfseries Initialize $x^i \sim p(x_{t+1}|x_t^i), i=1,2, \ldots, n$}
        \STATE {\bfseries Compute:} $\{\hat{x}^i\}=\text{SVGD}\left(p(x_{t+1}|Z_{t+1}),\{x^i \} \right) $.
        \STATE {\bfseries Output:  $\{x_{t+1}^i\}_{i=1}^n=\{\hat{x}^i\}_{i=1}^n$}
        \ENDIF
    \end{algorithmic}
\end{algorithm}

\subsection{Sliding window Stein particle filtering}
We next present a sliding window version of the Stein particle filtering algorithm. The main idea is to approximate the posterior distribution $p(x_{t+1:t+T} | Z_{t+T})$ over state trajectories of length $T$ recursively. When $t=0$, we sample from $p(x_1|Z_1)$ the same as Equation (\ref{eq:one_step_p1}).
When $1 \leq t < T$, the number of observations are not enough to form a window. By the Bayesian formula,
\begin{align}
     & p(x_{1:t+1} | Z_{t+1})      \nonumber                                                                                                 \\
    \propto
     & p(z_{t+1}|x_{t+1})   p(x_{1:t+1},Z_{t})   \nonumber                                                                                   \\
    \propto
     & p(z_{t+1}|x_{t+1}) p(x_{t+1}|x_{t})  p(x_{1:t}|Z_{t}) \nonumber                                                                       \\
    \propto
     & \left[\prod_{k=1}^{t+1} p(z_{k}|x_{k}) \right] \left[\prod_{k=2}^{t+1} p(x_{k}|x_{k-1}) \right] p(x_1) .\label{eq:window_less_than_T}
\end{align}

We sample $\{x^i_{1:t+1}\}_{i=1}^n$ from $p(x_{1:t+1} | Z_{t+1})$ by SVGD. Note that $x^i_{1:t+1} \in \mR^{(t+1)d}$ are the samples from joint distribution and we only take $\{x^i_{t+1} \}_{i=1}^n$ as the output amongst them.

When $t>T$, we can take advantage of the receding window, that is
\begin{align*}
     & p(x_{t+1:t+T} | Z_{t+T})                                             \\
    \propto
     & p(z_{t+T}|x_{t+T})   p(x_{t+1:t+T},Z_{t+T-1})                        \\
    \propto
     & p(z_{t+T}|x_{t+T}) p(x_{t+T}|x_{t+T-1})  p(x_{t+1:t+T-1}|Z_{t+T-1}). \\
\end{align*}
It follows that
\begin{align*}
     & p(x_{t+1:t+T} | Z_{t+T})                                                                     \\
    \propto
     & p(x_{t+1}|Z_{t+1}) \prod _{k=2}^T p(z_{t+k}|x_{t+k})   p(x_{t+k}|x_{t+k-1})                  \\
    \propto
     & \int_{x_t} p(x_t | Z_t)\prod_{k=1}^T \left(p(z_{t+k}|x_{t+k}) p(x_{t+k} | x_{t+k-1})\right).
\end{align*}
When $p(x_t | Z_t)$ is represented by particles $\{x_t^i\}_{i=1}^n$, the above can be evaluated efficiently and the Stein variational gradient descent \eqref{eq:stein} can be utilized to sample from $p(x_{t+1:t+T} | Z_{t+T})$. With particle-approximated distribution, we have

\begin{align}
     & p(x_{t+1:t+T} | Z_{t+T})  \nonumber                                                                     \\
    \propto
     & \frac{1}{n} \left[\sum_{i=1}^n p(x_t^i | Z_t) p(x_{t+1}|x_{t}^i) \right]   p(z_{t+1}|x_{t+1}) \nonumber \\
     & \times \prod_{k=2}^T p(z_{t+k}|x_{t+k}) p(x_{t+k} | x_{t+k-1}) . \label{eq:window_larger_than_T}
\end{align}

A good initialization can be realized by extending the samples from $p(x_{t:t+T-1} | Z_{t+T-1})$ by using the dynamics \eqref{eq:dyn1}. One potential advantage of this sliding window method is that it reuses some past observations to improve the current estimation and may correct some mistakes made in the past estimations.

\begin{algorithm}[tb]
    \caption{Sliding window Stein particle filtering}
    \label{al:window}
    \begin{algorithmic}
        \IF{$ t = 0$}
        \STATE {\bfseries Input:} initial dist. $p(x_1)$, observation $Z_1$
        \STATE { \bfseries Initialize $x^i \sim p(x_1), i=1,2, \ldots, n$}
        \STATE {\bfseries Compute:} $\{\hat{x}^i\}=\text{SVGD}\left(p(x_{1}|Z_{1}),\{x^i \} \right) $ with Algorithm \ref{al:svgd}.
        \STATE{\bfseries Output:}$ \{x_{1}^i\}_{i=1}^n=\{\hat{x}^i\}_{i=1}^n$
        \ELSIF{$1 \leq t < T$}
        \STATE {\bfseries Input:}
        particles $\{x_{1:t}^i\}_{i=1}^n \sim p(x_{1:t}|Z_t)$, observations $Z_t$
        \STATE { \bfseries Initialize $x^i \sim p(x_{t+1}|x_t^i), i=1,2, \ldots, n$}
        \STATE { \bfseries Concatenate $X^i = \left[x_{1}^i,\ldots x_{t}^i,x^i \right]$}
        \STATE {\bfseries Compute:} $\{\hat{X}^i\}=\text{SVGD}\left(p(x_{1:t+1}|Z_{t+1}),\{X^i \} \right) $.
        \STATE {\bfseries Output} $\{x_{t+1}^i\}_{i=1}^n$ which are the last dimension particles from $\left\{\widehat{X}^i \right\}$
        and then save $\{x_{1:t+1}^i\}_{i=1}^n=\left\{\widehat{X} \right\}$ for next time input
        \ELSIF{$t \geq T$}
        \STATE {\bfseries Input:}
        particles $\{x_{t-T+2:t}^i\}_{i=1}^n \sim p(x_{t-T+2:t}|Z_t)$ , observations $Z_t$
        \STATE { \bfseries Initialize $x^i \sim p(x_{t+1}|x_t^i), i=1,2, \ldots, n$}
        \STATE { \bfseries Concatenate $X^i = \left[x_{t-T+2}^i,\ldots x_{t}^i,x^i \right]$}
        \STATE {\bfseries Compute:} $\{\hat{X}^i\}=\text{SVGD}\left(p(x_{t-T+2:t+1}|Z_{t+1}),\{X^i \} \right) $.
        \STATE {\bfseries Output:} $\{x_{t+1}^i\}_{i=1}^n$ which are the last dimension particles from $\left\{\widehat{X}^i\right\}$
        and then save $\{x_{t-T+3:t+1}^i\}_{i=1}^n=\{\widehat{X}\}$
        \ENDIF
    \end{algorithmic}
\end{algorithm}



\section{Numerical examples} \label{sec:num}
{
In this section, we provide numerical examples to evaluate the performance of our proposed algorithm in comparison with the SIR particle filtering algorithm. In all the reported numerical experiments, the number of particles of our algorithm and SIR particle filter are the same, equal to $n = 500$.  For the SVGD Algorithm~\ref{al:svgd}, the iteration number $L=100$, and the step-size $\epsilon_\ell = 0.01$. 
}

\subsection{Linear Gaussian Setting}
Consider the following continuous-time filtering problem with linear dynamics, linear observation model, and Gaussian prior distribution:
\begin{subequations}
    \begin{eqnarray}
        dx_t &=& -\frac{1}{2} x_t dt+dW_t,\quad x_0 \sim \cN(1,1) \\
        dy_t &=& 3 x_t dt+ \frac{1}{2} dV_t, 
    \end{eqnarray}
\end{subequations}
where $\{W_t\}$, $\{V_t\}$ are mutually independent standard Wiener processes. The solution to this filtering problem is explicitly known, given by the Kalman-Bucy filter~\cite{kalman1961new}. The posterior distribution is Gaussian $\cN(\mu_t,\Sigma_t)$ with the update law for mean and covariance as follows:
\begin{subequations}
    \begin{align}
        d \mu_t &= -\frac{1}{2} \mu_t dt + K_t(d y_t - 3 \mu_t dt)\\
        \frac{d \Sigma_t}{d t}  &= - \Sigma_t + 1 - 36\Sigma_t^2
    \end{align}
\end{subequations}
where $K_t = 12\Sigma_t$ is the Kalman gain. 

In order to implement our proposed algorithm and the SIR particle filter, the  continuous-time system is discretized according to
\begin{subequations}
    \begin{align}
       x_{t+\Delta t} &= (1-\frac{1}{2}\Delta t )  x_t + W_{t+\Delta t} - W_t,\\
         z_t := \frac{y_{t+\Delta t} - y_{t}}{\Delta t}  &= 3 x_t  +\frac{1}{2\Delta t}(V_{t+\Delta t} - V_t) 
    \end{align}
\end{subequations}
where time step-size $\Delta t = 0.02$, and we introduced the discrete-time observation signal $z_t$. The increment of the Wiener process $W_{t+\Delta t} - W_t$ and $V_{t+\Delta t} - V_t$  is simulated by sampling independent  Gaussian random variables from $\cN(0,{\Delta t})$.

In order to measure the performance of the filtering algorithm, we use the mean-square-error criteria for estimating the conditional and the conditional covariance, averaged over $M=50$ independent runs: 
\begin{subequations}
\begin{align}
    \text{ m.s.e}_t^\mu   & =\frac{1}{M} \sum_{m=1}^M \left( \mu^{(n)}_t  - \mu_t \right)^2 , \label{mean mse}      \\
    \text{m.s.e.}_t^\Sigma & =\frac{1}{M} \sum_{m=1}^M \left(  \Sigma^{(n)}_t - \Sigma_t \right)^2, \label{cov mse}          
\end{align}
\end{subequations}
where $\mu^{(n)}_t$ and $\Sigma^{(n)}_t$ are empirical mean and covariance of the particles given by
\begin{subequations}
\begin{align}
    \mu^{(n)}_t       & =\frac{1}{n} \sum_{i=1}^n x^i_t,        \quad 
           \Sigma^{(n)}_t    & =\frac{1}{n-1} \sum_{i=1}^n (x_t^i-\mu^{(n)}_t)^2,\nonumber
\end{align}
\end{subequations}

\begin{figure}[h]
    \centering
    \begin{subfloat}[m.s.e. for mean \label{fig:linear_mean}]
        { \includegraphics[width=0.95\linewidth]{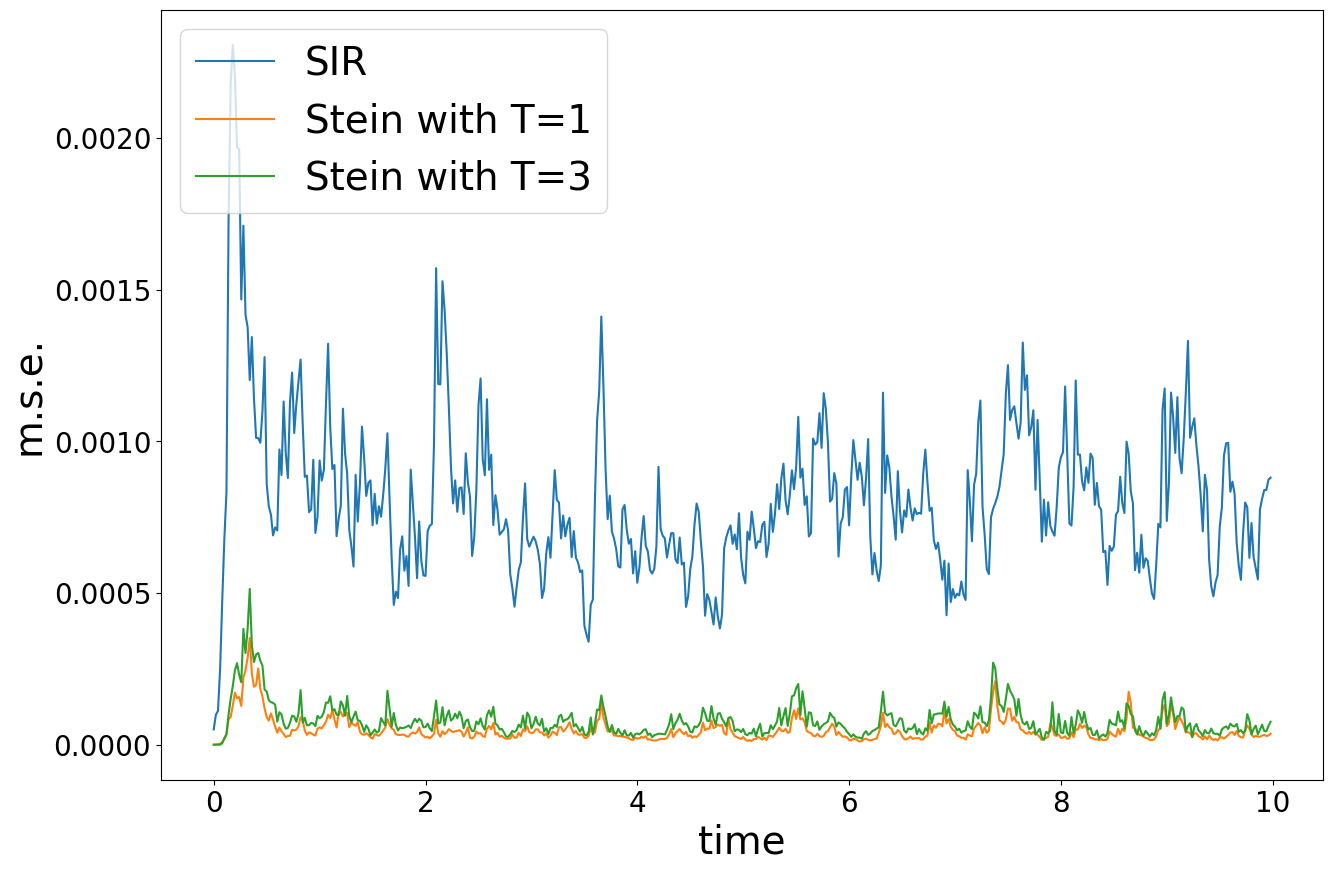}}
    \end{subfloat}

    \begin{subfloat}[m.s.e. for covariance \label{fig:linear_cov}]
        {\includegraphics[width=0.95\linewidth]{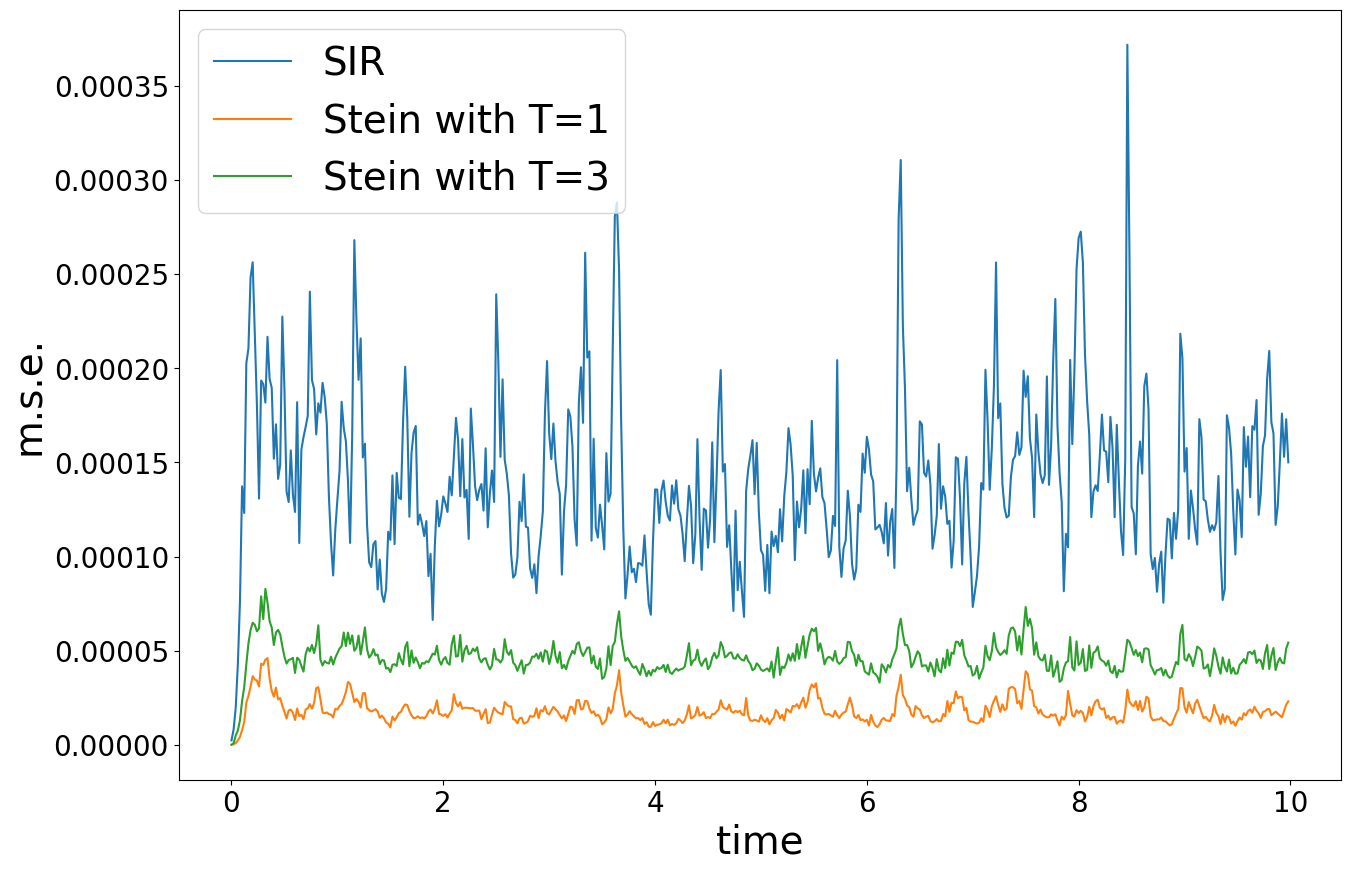}}
    \end{subfloat}
    \caption{Numerical comparison of the filtering algorithms: Stein method and SIR particle filter, for the Linear Gaussian example}
    \label{fig:linear}
\end{figure}

The numerical result of the m.s.e. error for our proposed algorithm with stride $T=1$ and $T=3$, and the SIR particle filter are depicted in Figure~\ref{fig:linear}. It is observed that our proposed algorithm  admits smaller error compared to SIR with either choice of window length $T=1,3$. Among the two choices for time window, the one with $T=1$ admits slightly smaller error which is probably due to the fact that the particles in this case belong to $\mathbb{R}$, while in $T=3$, the particles belong to $\mathbb{R}^3$. We conjecture the optimization in a higher dimensional space could jeopardize the performance.

\subsection{Nonlinear Case}
Consider the Benes filtering problem
\begin{subequations}
    \begin{eqnarray}
        dx_{t} &=& \mu \sigma_B \tanh\left(\frac{\mu}{\sigma_B }x_t \right) dt+\sigma_B dW_t, \\
        dy_t &=& (h_1x_t +h_1h_2)dt +dV_t,
    \end{eqnarray}
\end{subequations}
with $\mu=0.1$, $\sigma_B=0.3$, $h_1=5$, and $h_2=0$.  

The closed-form solution of Benes filter is a mixture of two Gaussians~\cite[Ch. 6]{bain2008fundamentals}:
\begin{align*}
    c_t \cN(a_t-b_t,\sigma_t^2)+(1-c_t)\cN(a_t+b_t,\sigma_t^2),
\end{align*}
where
\begin{align*}
    a_t        & =  \sigma_B \Psi_t \tanh(h_1 \sigma_B t)+\frac{h_2+x_0}{\cosh(h_1 \sigma_B t)} -h_2, \\
    b_t        & =  \frac{\mu}{h_1} \tanh(h_1 \sigma_B t),                                            \\
    \sigma_t^2 & =  \frac{\sigma_B}{h_1} \tanh(h_1 \sigma_B t) ,                                      \\
    \Psi_t^2   & = \int_0^t \frac{\sinh(h_1 \sigma_B s)}{\sinh(h_1 \sigma_B t)} dY_s ,                \\
    c_t        & =  \frac{1}{1+e^{\frac{2a_tb_t}{\sigma_B} \coth(h_1 \sigma_B t)}}.
\end{align*}

\begin{figure}[h]
    \centering
    \begin{subfloat}[posterior density \label{fig:benes_density}]
        {\includegraphics[width=0.95\linewidth]{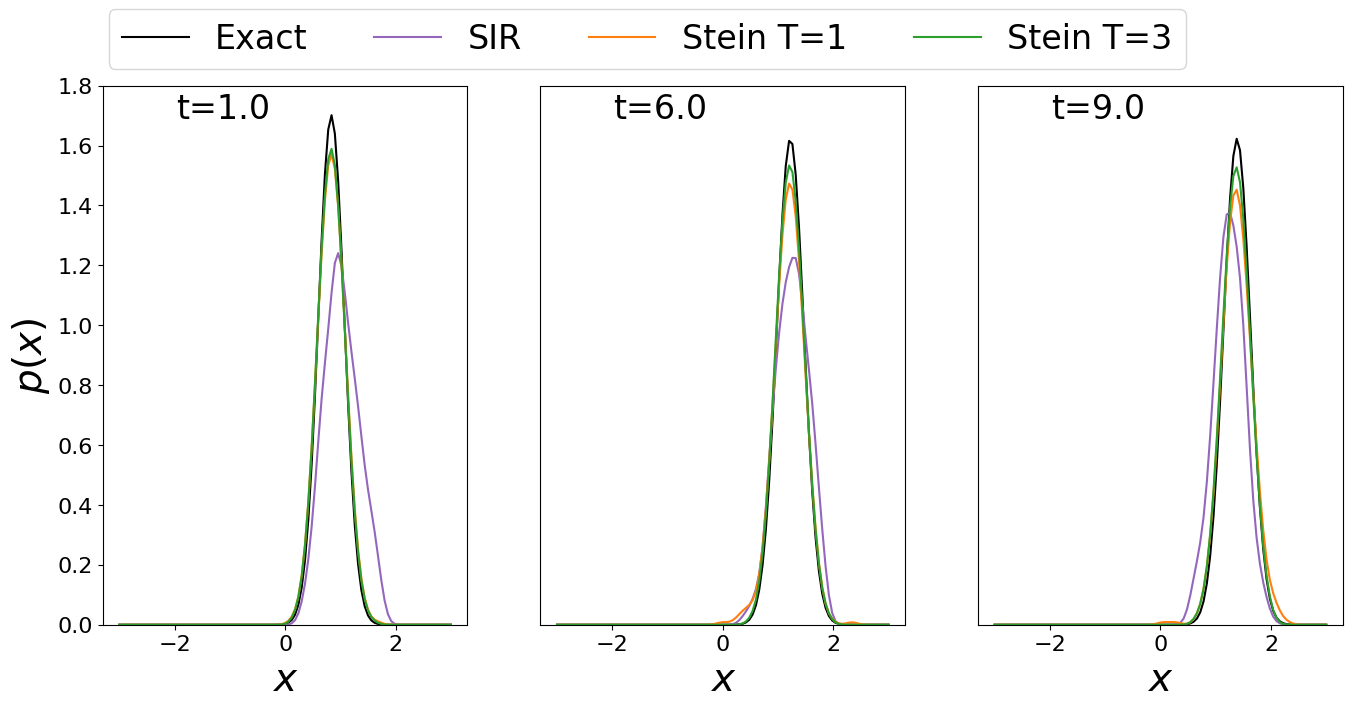}}
    \end{subfloat}
    
    \begin{subfloat}[m.s.e. of the mean \label{fig:benes_mean}]
        { \includegraphics[width=0.95\linewidth]{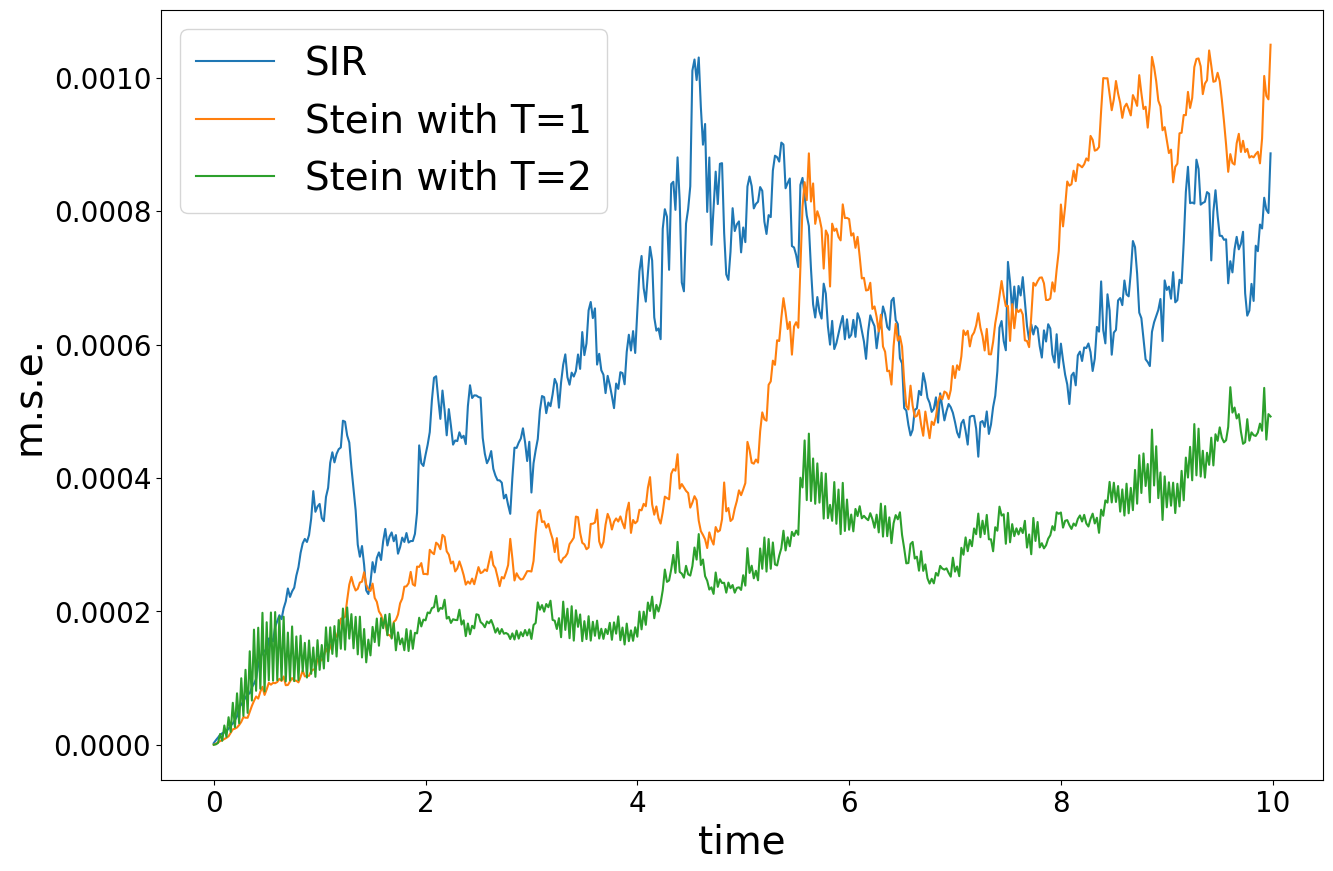}}
    \end{subfloat}
    \caption{Numerical comparison of the filtering algorithms, Stein method and SIR particle filter, for the Benes filter. }
    \label{fig:benes}
\end{figure}

In order to implement our algorithm and the SIR particle filter, we discretize the continuous-time model similar to the linear Gaussian case  with time step-size $\Delta t = 0.02$.  The numerical results for the Benes filter is depicted in Figure~\ref{fig:benes}. The sub-figure~\ref{fig:benes_density} depicts the exact posterior distribution along with the distributions of particles from the filtering algorithms at three time instants. In order to evaluate the density from particles, we used a kernel density estimation (KDE) with Gaussian kernel with bandwidth $0.1$.  The sub-figure~\ref{fig:benes_mean} depicts the m.s.e. error of the mean defined according to Equation~\eqref{mean mse} averaged over $M=50$ independent runs. It is observed that the Stein method with time window $T=3$ admits smaller error. This example demonstrates the advantage of using a time-window. It seems that in this case, the advantage of processing more observation signal  surpasses the error brought by simulating over a higher dimensional space.  

\section{Conclusion}\label{sec:conclusion}
We propose a novel particle filtering framework based on the Stein method for variational inference. Two different implementations of this Stein particle filtering method are presented: sequential Stein particle filtering and sliding window Stein particle filtering. The former is a natural adoption of SVGD following the Bayes rule in particle filtering problems.
The latter makes use of multiple observations for calculating the posterior over a fixed-size time window.
How to choose the window length in the sliding window method may depend on a few factors, including the variance ratio between measurement noise and the state, and the error introduced by SVGD solver which grows with the window size. Both of the methods have shown advantages over SIR in the given examples.

One downside of the proposed Stein particle filter is that it is more computationally demanding than SIR due to additional iterations in the SVGD algorithm. Besides, the Stein particle filter becomes less effective when the variance of observation noise is dramatically larger than the variance of state. Empirically, we observed that the proposed method become more effectively when the randomness in the dynamics is relatively high and the measurement noise is relatively low. Note that this is known to be a challenging scenario for most particle filtering algorithms \cite{GriStaBur07}.

	{
		\bibliographystyle{IEEEtran}
		\bibliography{./refs}
	}
\end{document}